# Manuscript Details

| | |
|---|---|
| **Manuscript number** | JCOMB_2018_501 |
| **Title** | Performance-driven 3D printing of continuous curved carbon fibre reinforced polymer composites: a preliminary numerical study |
| **Article type** | Full Length Article |

**Abstract**

This paper presents a new concept to place continuous curved fibres for carbon fibre reinforced polymer (CFRP) composites, which can be fulfilled by potential additive or hybrid manufacturing technology. Based on the loading condition, principal stress trajectories are generated through finite element analysis (FEA) and used as the guidance for the placement paths of carbon fibres. Three numerical cases including an open-hole single ply lamina under uniaxial tension and open-hole cross-ply laminate under biaxial tension and normal pressure are studied and compared with traditional reinforced composites with unidirectional fibres. The modelling results show that the stress concentration in both fibre and matrix are reduced significantly by the curved fibre placement and the stiffness of CFRP composites have been improved. This concept of performance-driven optimization method could lead to a useful tool for the design of future 3D printing process for fibre reinforced composites.

| | |
|---|---|
| **Keywords** | CFRP; Curved fibres; Performance-driven manufacturing; 3D printing; FEA. |
| **Manuscript region of origin** | Europe |
| **Corresponding Author** | Dongmin Yang |
| **Corresponding Author's Institution** | University of Leeds |
| **Order of Authors** | Haoqi Zhang, Dongmin Yang, Y Sheng |
| **Suggested reviewers** | Badis Haddag, Daxu Zhang, Dianzi Liu |

## Submission Files Included in this PDF

**File Name  [File Type]**

cover letter.docx  [Cover Letter]

Performance-driven 3D printing of continuous curved CFRP_submission.pdf  [Manuscript File]

To view all the submission files, including those not included in the PDF, click on the manuscript title on your EVISE Homepage, then click 'Download zip file'.

Dear Editors:

We are pleased to submit our manuscript entitled "Performance-driven 3D printing of continuous curved fibre reinforced polymer composites: a preliminary numerical study" to Composites Part B: Engineering.

In this manuscript we presents a new concept to place continuous curved fibres along principal stress trajectories based on the loading condition for CFRP composites, which can be fulfilled by potential additive or hybrid manufacturing technology. We carried out three numerical cases to demonstrate its advantages, i.e. an open-hole single ply lamina under uniaxial tension and an open-hole cross-ply laminate under biaxial tension and normal pressure. When compared with traditional reinforced composites with unidirectional fibres, we found that the stress concentration in both fibre and matrix are reduced significantly by the curved fibre placement and the stiffness of CFRP composites have been improved. This concept of performance-driven optimization method could lead to a useful tool for the design of future 3D printing process for fibre reinforced composites.

Best regards,

Dr Dongmin Yang (corresponding author)

# Performance-driven 3D printing of continuous curved carbon fibre reinforced polymer composites: a preliminary numerical study

Haoqi Zhang, Dongmin Yang[*], Yong Sheng

School of Civil Engineering, University of Leeds, Leeds LS2 9JT, UK

**Abstract**

This paper presents a new concept to place continuous curved fibres for carbon fibre reinforced polymer (CFRP) composites, which can be fulfilled by potential additive or hybrid manufacturing technology. Based on the loading condition, principal stress trajectories are generated through finite element analysis (FEA) and used as the guidance for the placement paths of carbon fibres. Three numerical cases including an open-hole single ply lamina under uniaxial tension and open-hole cross-ply laminate under biaxial tension and normal pressure are studied and compared with traditional reinforced composites with unidirectional fibres. The modelling results show that the stress concentration in both fibre and matrix are reduced significantly by the curved fibre placement and the stiffness of CFRP composites have been improved. This concept of performance-driven optimization method could lead to a useful tool for the design of future 3D printing process for fibre reinforced composites.

---

[*] Corresponding author. E-mail address: d.yang@leeds.ac.uk (D. Yang)

# 1. Introduction

Carbon fibre reinforced polymer (CFRP) composites are widely used in aerospace, automobile and infrastructure industry because of its high strength and high stiffness-to-weight ratio. In the 1960s, the Automated Fiber Placement (AFP) technology was developed for the manufacturing of large structures by placing multiple layers of composite lamina to form a laminate. Input materials for the placement process are mainly unidirectional fibers, which are pre-impregnated (prepreg) with the matrix material [1]. Compared with AFP, 3D printing (also known as additive manufacturing) is a process that enables the efficient manufacturing of parts with complex shapes. Instead of using prepreg as in AFP, 3D printing uses filament or bundle of fibres and produces small size of printing beads to form a part [2] without the need of molds. Thus it is able to manufacture composites at low labor and production costs with an immense range of complexities [3].

Recently, a few new 3D printing technologies for CFRP composites have been developed. Among them, fused deposition modelling (FDM®) [4] developed by Stratasys Ltd. is able to print continuous aligned carbon fibres as shown in Fig. *1* [5]. The mechanical performance of these composites under different loading condition have been evaluated [6-12]. On the other hand, new design principles and methods in printing progress are being developed, such as hot-press treatment to reduce porosity and improve the strength and stiffness of the printed composites [13-17].

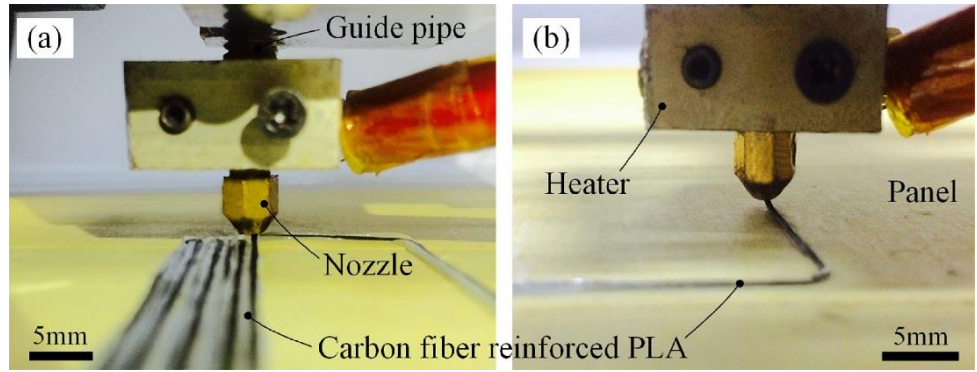

Fig. 1. 3D printing process of the continuous carbon fiber reinforced composite [5]: (a) Printing the straight area of composite sample (b) Printing the corner of composite sample.

For composite laminate, the stacking sequence of each ply with different angles must be optimised to achieve the designed stiffness on each direction [18-22]. However, the fibres are placed unidirectionally in each ply of the laminate and drilling machines are used to manufacture the notched composite specimens in the traditional manufacturing process [23-26]. Thus, the outstanding properties of fibres may be underutilized and most of fibres near the hole are cut off, which provokes severe stress concentration and reduces stiffness and strength of constructions with geometric discontinuities.

The new additive manufacturing technology offers the opportunity to change the orientation of fibres during the manufacturing process, therefore, an optimization method of continuous fibre placement is required for the 3D printed CFRP composites. To fully utilize the properties of fibre, a few algorithms have been developed to adjust the placement of fibres [27, 28]. These methods usually require lots of variables and some assumptions, which complicates the process of optimization. In addition, the orientation of CFRP material has to be optimised in each element when carrying out finite element analysis, without considering whether the real manufacturing process is possible to achieve it or not [28-30]. It also brings discrete fibre angles in the mesh and large discrepancy of fibre angles between the adjacent elements. Also, the stress distribution is inaccurate since CFRP lamina is treated as an equivalent homogeneous material in these numerical simulation [27].

To maximise the potential of 3D printing for lightweight CFRP composite structure, a feasible method for continuous fibre placement is essentially required. For composites, material with excellent tensile property and slender shape is usually placed along principal stress trajectories to improve the mechanical performance, such as steel bars in the concrete. Also, some researchers tried to make cellular materials with cell walls that align with the stress trajectories in order to improve load distribution [31]. In this paper, we present a new concept of printing long carbon fibre along the trajectories of maximum or middle principal stress. Finite element analysis of specimens with only matrix material is first used to generate the trajectories of principal stress. We carry out numerical studies on three cases including single ply lamina under uniaxial tension and cross-ply laminates under biaxial tension and normal pressure, where fibres and matrix are defined as different isotropic materials. The change of stress distribution and stiffness of models are discussed and compared to traditional composite lamina or laminate.

## 2. Methodology

Since the tensile properties of carbon fibres are superior in the fibre direction, we aim to place fibres along the direction of maximum tensile stresses. In other words, the optimised path of fibres should align with stress trajectories which are lines whose direction at each point gives the orientation of the maximum or middle principal stress. For the optimization problem, we first simulate the specimen printed using only matrix material under uniaxial tension. With the data of in-plane stresses of each node from finite element analysis, we calculate the orientations and values of maximum principal stresses by Eqs. 1-2.

$$\tan \varphi = -\frac{\sigma_x-\sigma_y}{2\tau_{xy}} \pm \sqrt{1+(\frac{\sigma_x-\sigma_y}{2\tau_{xy}})^2} \qquad (1)$$

$$\sigma_1/\sigma_2 = \frac{\sigma_x+\sigma_y}{2} \pm \sqrt{(\frac{\sigma_x-\sigma_y}{2})^2 + \tau_{xy}^2} \qquad (2)$$

where $\sigma_x$ is stress in x-axis direction, $\sigma_y$ is stress in y-axis direction, $\tau_{xy}$ is shearing stress, $\varphi$ is the angle between the stress normal to the stress trajectories at a point and the x-axis, $\sigma_1$ is the value of maximum principal stress, $\sigma_2$ is the value of middle principal stress (Two values of angle $\varphi$ are calculated for maximum and middle principal stress). Then the coordinates and angles for maximum/middle principal stress of each nodes are imported into Tecplot 360 software. Stress trajectories as streamlines are created by entering start position, gap and number of streamlines. We can also change the number of the streamlines in a ply in order to obtain models with different volume fraction of fibres. For case studies in this paper, these streamlines are extracted and imported into a FEM software (ABAQUS) with a specific width as a part, therefore, optimised models with curved fibres can be studied.

As shown in Fig. *2*, Case One focuses on modelling an open-hole single ply lamina,

where its length is 100 mm and its width is 50 mm. The radius of the hole is 20 mm and the thickness of the plate is 10mm. The tensile stress of 40 MPa is applied to the right side of the plate. The left side of the plate has a constrained degree of freedom in horizontal displacement along the x-axis. The stress distribution is obtained by FEA with 5748 quadrilateral element and 5844 nodes. Mechanical properties of carbon fibres and matrix used in present study are listed in Table 1. As shown in Fig. *3*, eight maximum principal stress trajectories are generated for the printing path of fibres ($V_f = 3.4\%$).

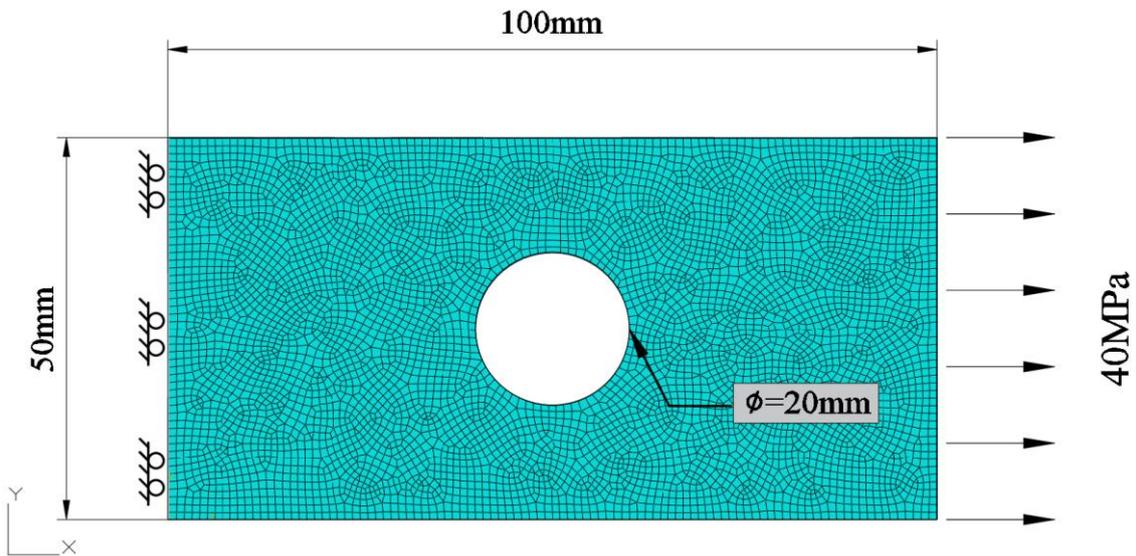

Fig. 2. Configuration of the CFRP plate with a hole under uniaxial tension (Case One).

Table 1 Mechanical properties of carbon fibre and epoxy [32].

| Material | Modulus of elasticity | Poisson's ratio |
| --- | --- | --- |
| IM7 Carbon fibres | 276 GPa | 0.28 |
| Hexcel HexPly 8552 Epoxy matrix | 4.76 GPa | 0.37 |

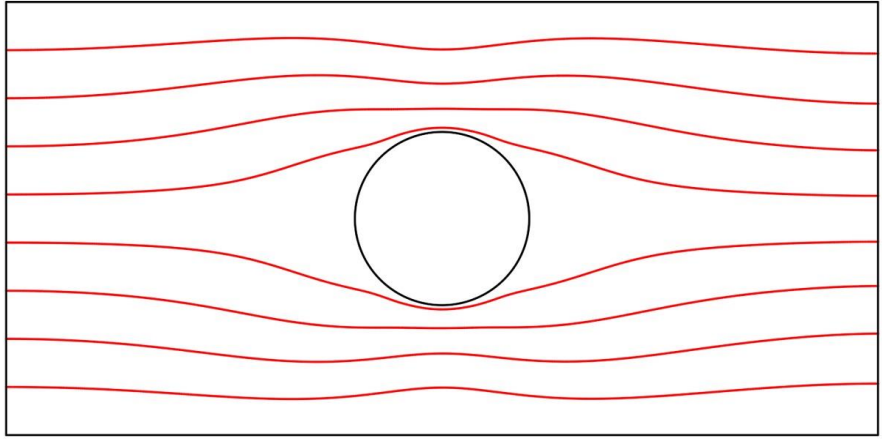

Fig. 3. Generated maximum principal stress trajectories (red lines) at $V_f = 3.4\%$.

In this optimization method, fibres are simulated as a bundle with the width of 0.2 mm (The print resolution of present commercial 3D printers can be 100 microns, such as Mark Two Printer supplied by MarkForged®, Somerville, MA) along the generated stress trajectories as shown in Fig. *4*. In this specific case, an equivalent volume fraction of fibres $V_f$ between 1.7% and 27.2% without any overlap of fibres can be achieved. For the conventional optimised composites, fibres are placed the along x-axis direction straightly as shown in Fig. *5*.

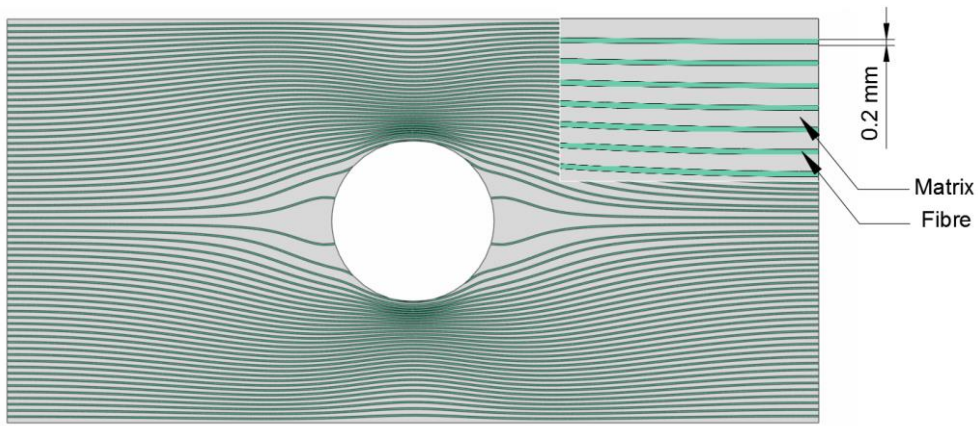

Fig. 4. Optimised CFRP plate with an equivalent volume fraction of fibres $V_f = 27.2\%$.

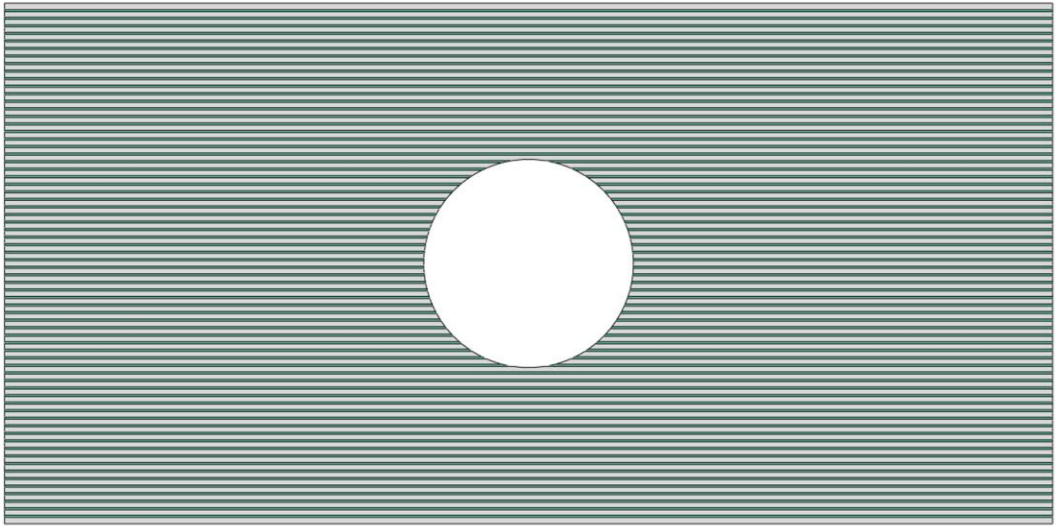

Fig. 5. FEM model of a conventionally optimised CFRP plate with an equivalent volume fraction of fibre $V_f = 27.2\%$.

## 3. Numerical case studies

Three cases are investigated in this paper. Besides Case One shown above, Case two is an open-hole laminate under biaxial tension with the same length, width and radius of the hole. As shown in Fig. *6*(a), the tensile stress of 40 MPa is applied to the right and bottom sides of the laminate. The left side of the laminate has a constrained degree of freedom in horizontal displacement along the x-axis while top side has a constrained degree of freedom in vertical displacement along the y-axis. The optimised model consists of 4 piles and the thickness of each layer is 2 mm. The biaxial tension is considered as the combination of two perpendicular uniaxial tension. We optimize the fibre alignment in each ply, therefore, the stress trajectories in each layer can be obtained from the simulation under uniaxial tension as shown in Fig. *6*(b) and Fig. *6*(c). With the same stacking sequence and thickness, conventionally optimised CFRP plate with a hole consists of 4 plies where fibres are placed along 0 and 90 degrees symmetrically, i.e. (0/90)s.

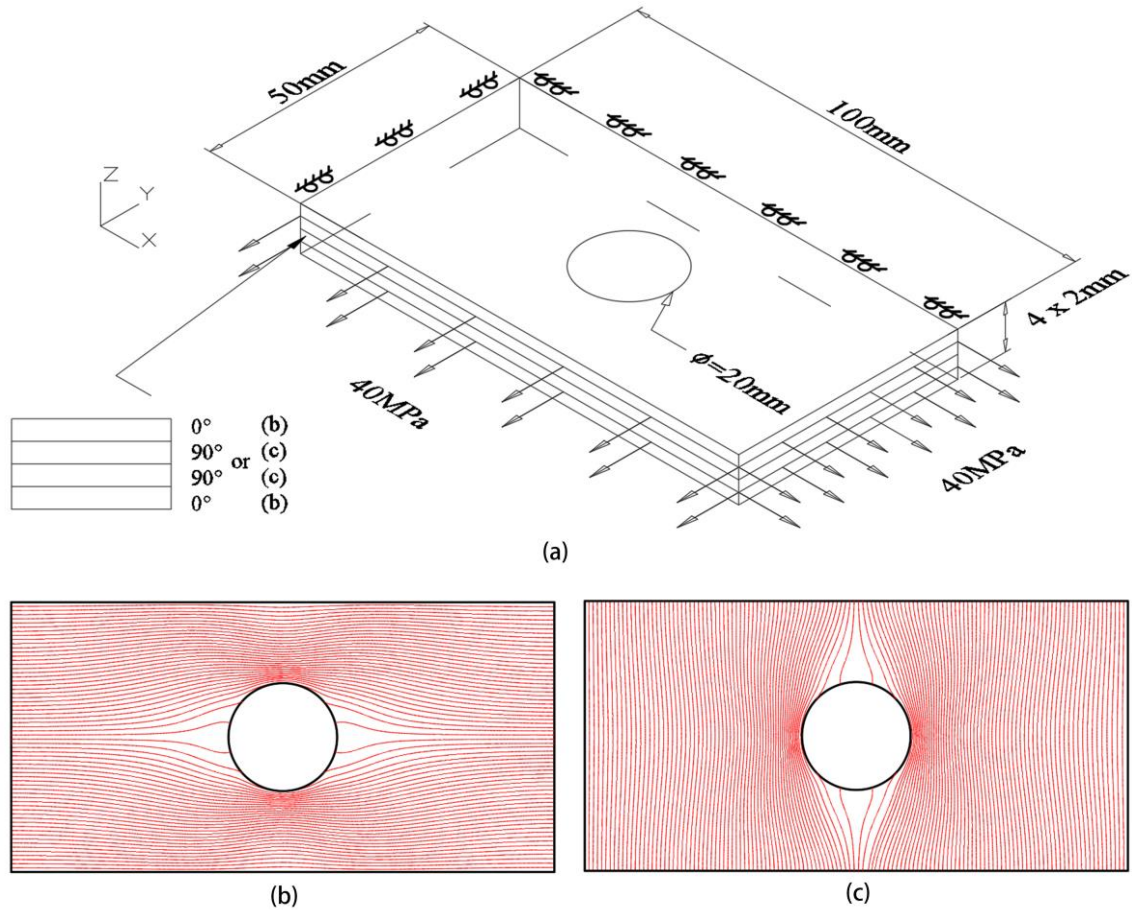

Fig. 6. Optimised CFEP plate under biaxial tensile loading (Case Two, $V_f = 27.2\%$): (a) Configuration and stacking sequence (b) Stress trajectories generated under uniaxial tension along x-axis direction (c) Stress trajectories generated under uniaxial tension along y-axis direction

Case Three is an open-hole laminate under normal pressure. The size of laminate as well as the number and thickness of plies are the same as in Case Two. As shown in *Fig. 7*(a), the normal pressure of 40 MPa is applied to the top face of laminate and its four sides are fixed. As the pressure is applied in the out-of-plane direction of the plate and consider the symmetric layout of the laminate, for optimised fibres placement in plies, two kinds of optimization paths are applied, where fibres are placed along maximum and middle principal stress trajectories as shown in *Fig.* 7(b) and *Fig.* 7(c). The stacking sequence is shown in Figure 8, where plies are set symmetrically, i.e. (Maximum/Middle)s. In order to compare the improvement of mechanical performance, two different traditionally optimised models with unidirectional fibres are simulated, i.e. (-45°/45°)s and (90°/0°)s. Because of the symmetrical geometry, material and boundary conditions of Case Three, we only simulate a quarter of the plate in the following numerical study for this case.

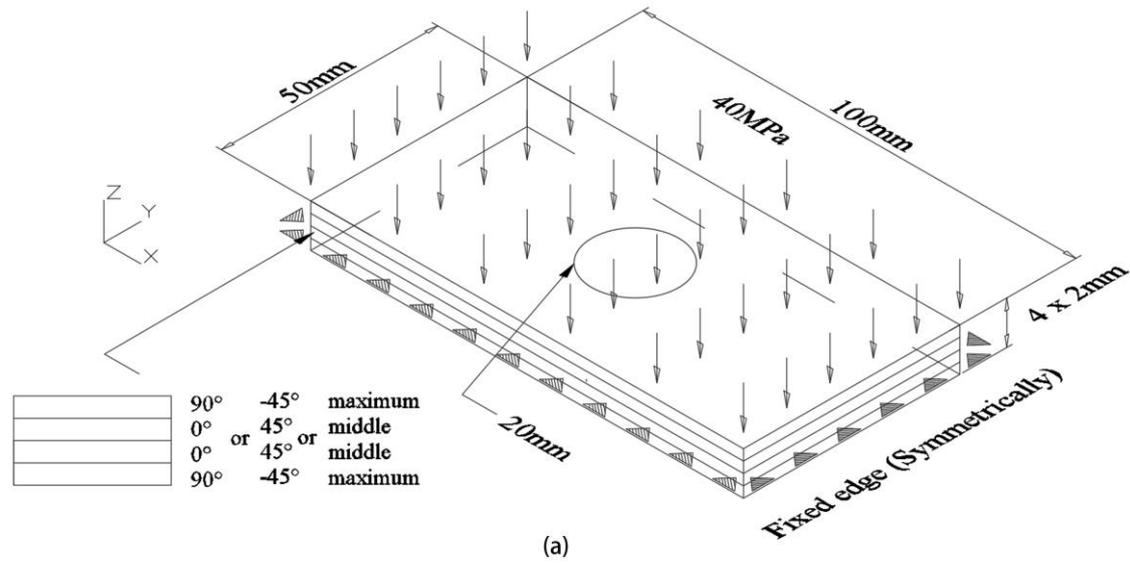

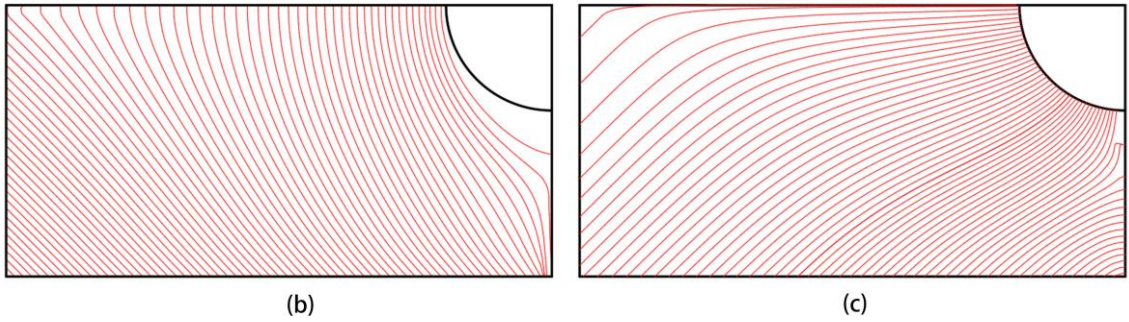

Fig. 7. Optimised CFEP plate under normal pressure (Case Three, $V_f$ =27.1%): (a) Configuration and stacking sequence (b) Maximum principal stress trajectories (c) Middle principal stress trajectories.

## 4. Results and discussion

FEM models for all above mentioned cases were developed and numerical results were compared with those from models of composites with unidirectional fibre placements.

With Eqs. 3-5, the maximum von Mises stress in the regions of fibre and matrix respectively, the stiffness and the maximum stress concentration factor are calculated.

$$\sigma_s = \sqrt{\frac{(\sigma_1-\sigma_2)^2+(\sigma_2-\sigma_3)^2+(\sigma_1-\sigma_3)^2}{2}} \quad (3)$$

$$K_{tmax} = \frac{\sigma_{s-max}}{\sigma_0} \tag{4}$$

$$k = \frac{P}{\delta_{max}} \tag{5}$$

where $\sigma_1$ is the value of maximum principal stress, $\sigma_2$ is the value of median principal stress, $\sigma_3$ is the value of minimum principal stress, $\sigma_s$ is the von Mises stress, $\sigma_{s-max}$ is the maximum von Mises stress, $\sigma_0$ is the nominal stress, $K_{tmax}$ is the maximum stress concentration factor, $P$ is the force of the edge, $\delta_{max}$ is the maximum displacement, and $k$ is the stiffness of the composites. We discuss the stiffness of all these models and the maximum stress concentration factor in fibre and matrix of the first two series of models, since CFRP lamina is treated as an equivalent homogeneous material in FEA and the distribution of stress is not comparable with the first two heterogeneous models.

To evaluate whether the stress distribution in the locations adjacent to the hole is even or not, the average values of Von Mises stress in the matrix within a distance of 2 mm to the edge of the hole are calculated and plotted. As shown in *Fig. 8* (d), the values are represented by the distance between the points in the curve and the origin of the coordinate plane. Therefore, if the curve is similar to a circle with a small radius, stresses are small and also distribute evenly in the vicinity of the hole in this specific model.

For Case One, results from models without any overlap of fibres are represented by solid lines (the volume fraction of fibre between 1.7% and 27.2%) as shown in Fig. *8*. The maximum stress concentration factor and stiffness have little differences between new and traditional optimization method with the volume fraction of fibre 1.7%. As the fibre volume fraction increases, models with curved fibre show increasingly better performance. With the fibre volume fraction 27.2% (which is highest fibre volume fraction without any overlap of fibres), the maximum stress concentration factors in both fibre and matrix reduce by almost 60% (60.3% and 59.8% respectively). The stiffness in x-axis direction of the plate with curved fibres increases by 39.7% in comparison with the traditionally optimised models with unidirectional fibres. In terms of comparison between models with unidirectional fibres and model using lamina material, the stiffness of composites has a marginal error of 6.9%, which is acceptable and also indicates that the method modelling fibre and matrix separately is applicable.

Considering the improvement of printing technology in the future, we also simulate new optimised models with higher volume fraction of fibre, where the width of carbon fibre bundles are 0.3, 0.4 and 0.5 mm and the number of carbon fibre bundle is the same as in the model with a fibre volume fraction of 27.2%. Fibres are assumed to be printed without matrix material in the area near to the hole and the results are plotted out by dotted lines, as shown in Fig. *8*. The maximum stress concentration factor in the fibres decreases continuously and factor in the matrix fluctuates slightly when the volume fraction of fibre increases from 27.2% to 67.9%. Also the stiffness increases continuously. In terms of the comparison between models with curved and unidirectional fibres, new optimised models with fraction between 30% and 40% present better performance. Except for slightly fluctuating concentration factor in the matrix, factor in the fibre reduces by 60% and stiffness of composites increases by 57% approximately.

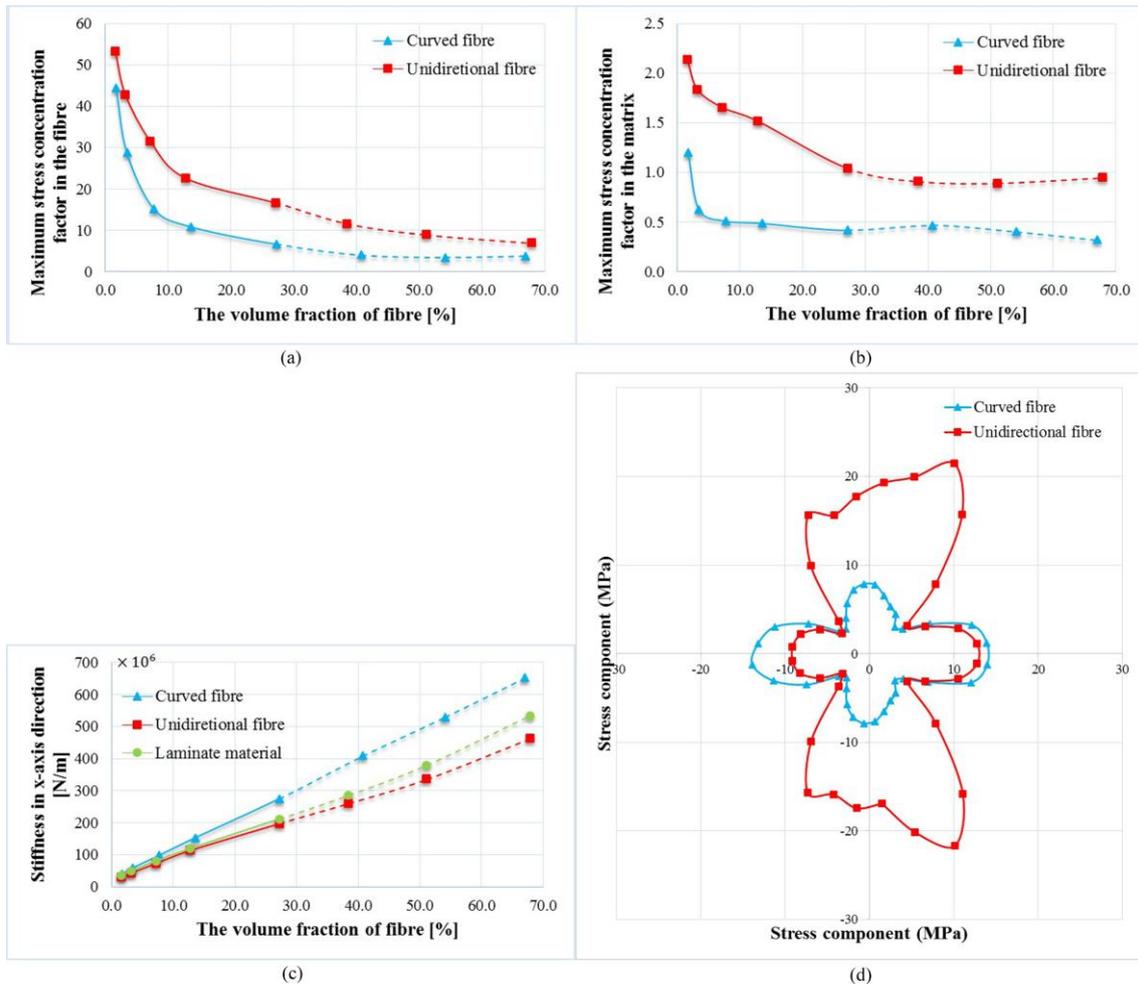

Fig. 8. Comparisons of mechanical performance in Case One: (a) maximum stress concentration factor in the fibre (b) maximum stress concentration factor in the matrix (c) stiffness in x-axis direction (d) Von Mises stress distribution in the matrix at the circumference of hole ($V_f = 27.2\%$)

In Table 2, the distribution field of Mises stress for Case One with the fibre volume fraction 27.2% are shown. For models with unidirectional fibres, the maximum Mises stresses, no matter in the fibres or matrix, concentrate in a very small area near to the top and bottom edge of the hole. It can also be observed that the Mises stresses are distributed more evenly in the models with curved fibres. The maximum stresses in the fibre distribute on a larger area adjacent to the hole, because of that, the maximum Mises stresses in the matrix do not concentrate on the top and bottom edge of the hole. But as shown in Fig. *8*(d), in the region adjacent to the hole, although the stresses reduce significantly, they do not distribute evenly since only carbon fibres in x-axis direction are optimised in this case with single ply.

Table 2 Von Mises stress distribution in Case One with a fibre volume fraction 27.2%

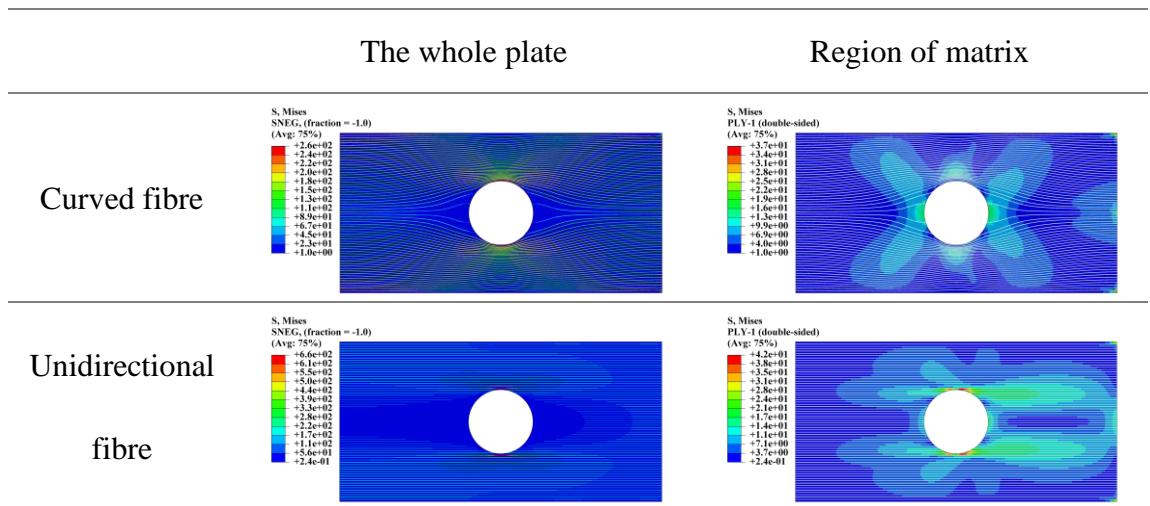

For Case Two, as shown in Fig. *9*, models with fibre volume fractions from 6.4% to 27.2% are simulated, which is a representative fraction range according to the result from Case One. Compared with traditional optimization method, models with curved fibres show better mechanical performance as the fraction increases. For the model with fraction 27.2%, the maximum stress concentration factor in the fibre and matrix reduces by 33.0% and 62.5% respectively. Since Case Two is an open-hole laminate under biaxial tension, the stiffness of the composites in both x-axis and y-axis direction is analysed, which increases by 24.6% and 49.3%.

Distribution field of Mises stress for Case Two is shown in Table 3. We consider two different types of fibre placement in a ply, x-axis direction layer and y-direction layer, which are compared with 0° and 90° plies in the traditional optimization method with unidirectional fibres. As can be seen in the comparison of stress distribution for the whole plate, the maximum stresses in the fibre distribute in a larger area around the hole and also have a smaller value. For the region of matrix, the Mises stresses distribute quite evenly and stress concentration is not obvious in the model with curved fibres while the Mises stress much concentrate near to the hole in the model with unidirectional fibres. Similarly, as shown in Fig. *9*(e) and Fig. *9*(f), Von Mises stresses with smaller values distribute evenly in the region adjacent to the hole, which proves that the new optimization method can reduce the stress concentration significantly in this case.

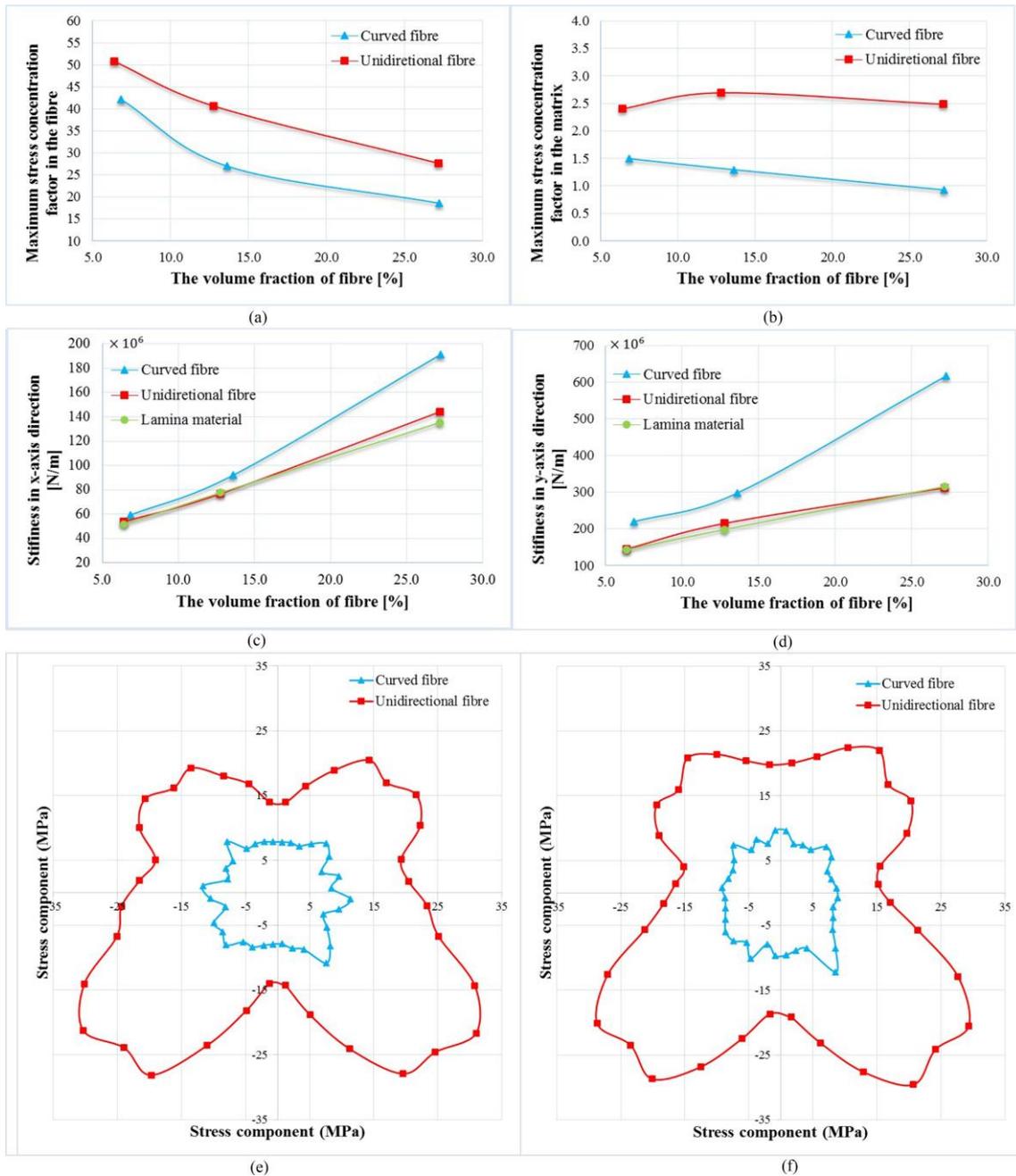

Fig. 9. Comparisons of mechanical performance in Case Two: (a) maximum stress concentration factor in the fibre (b) maximum stress concentration factor in the matrix (c) stiffness in x-axis direction (d) stiffness in y-axis direction (e) Von Mises stress distribution at the circumference of hole in x-axis direction ($V_f = 27.2\%$) (f) Von Mises stress distribution at the circumference of hole in y-axis direction ($V_f = 27.2\%$)

Table 3 Von Mises stress distribution in Case Two with a fibre volume fraction 27.2%

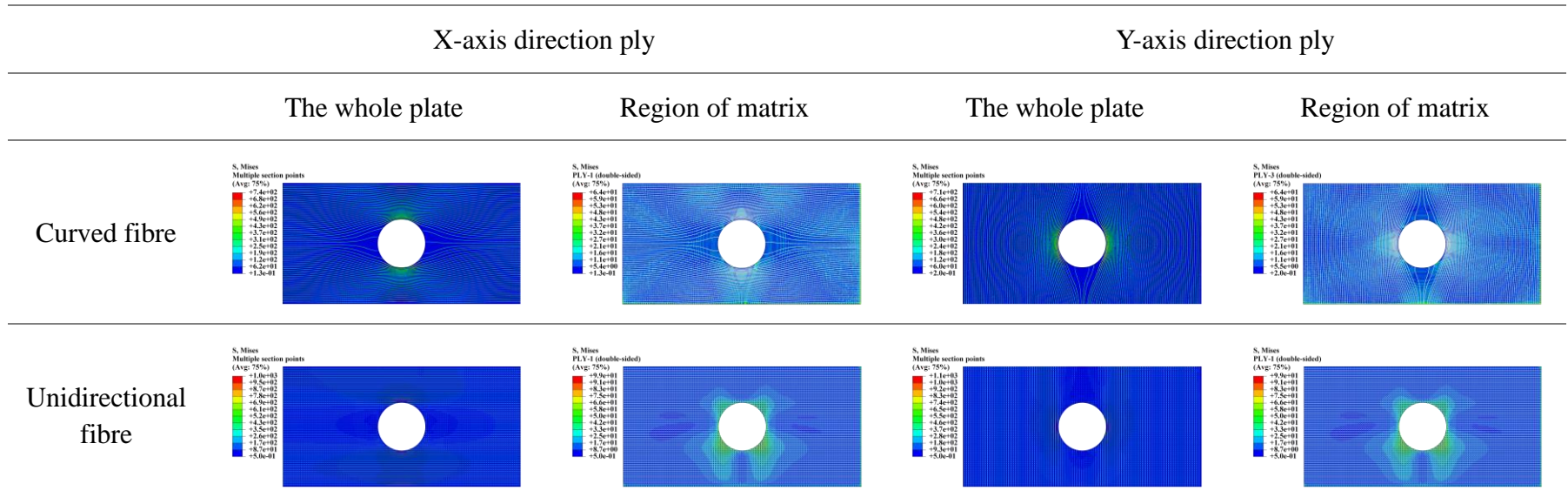

For Case Three, as shown in *Fig. 10*, models with stacking sequence (-45°/45°)s show better performance than models with stacking sequence (90°/0°)s both in maximum stress factor and stiffness. So we compare models with optimised curved fibres with the unidirectional models with fibre volume fraction from 16% to 36%. As in Case One and Two, the effect of optimization gradually diminishes when the overlap of fibres appear. Mechanical performance are improved significantly with a fibre volume fraction around 27.1%. The maximum stress concentration factor reduces by 28.9% and 29.0% in the fibre and matrix. Also the stiffness in z-axis direction increases by 29.7%.

As shown in Table 4, Von Mises stresses in the optimised model are well-distributed both in the fibre and matrix compared with the other two models. In the fibre, stresses with maximum values distribute on a wider region which is no longer near the hole. For the matrix, as shown in *Fig. 10*(d) and *Fig. 10*(e), the curves of Von Mises stress adjacent to the hole are more similar to a circle and the values of them are reduced significantly, which shows the new optimization method has a great impact on improvement of mechanical performance of laminate under normal pressure.

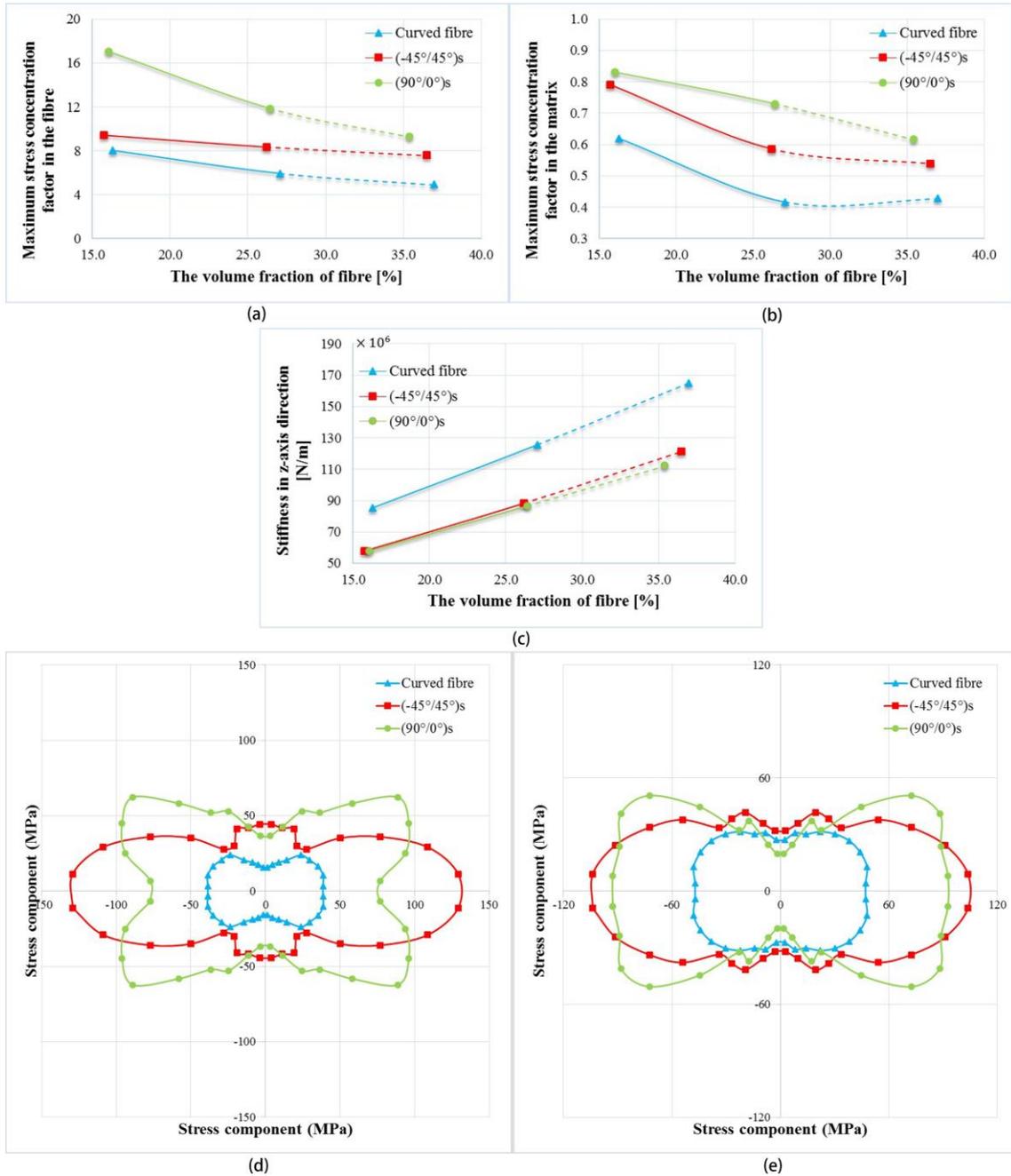

Fig. 10. Comparisons of mechanical performance in Case Three: (a) maximum stress concentration factor in the fibre (b) maximum stress concentration factor in the matrix (c) stiffness in z-axis direction (d) Von Mises stress distribution in the matrix at the circumference of hole in the top/bottom plies ($V_f = 27.1\%$) (e) Von Mises stress distribution at the circumference of hole in the middle two plies ($V_f = 27.1\%$)

Table 4 Von Mises stress distribution in Case Three with a fibre volume fraction 27%

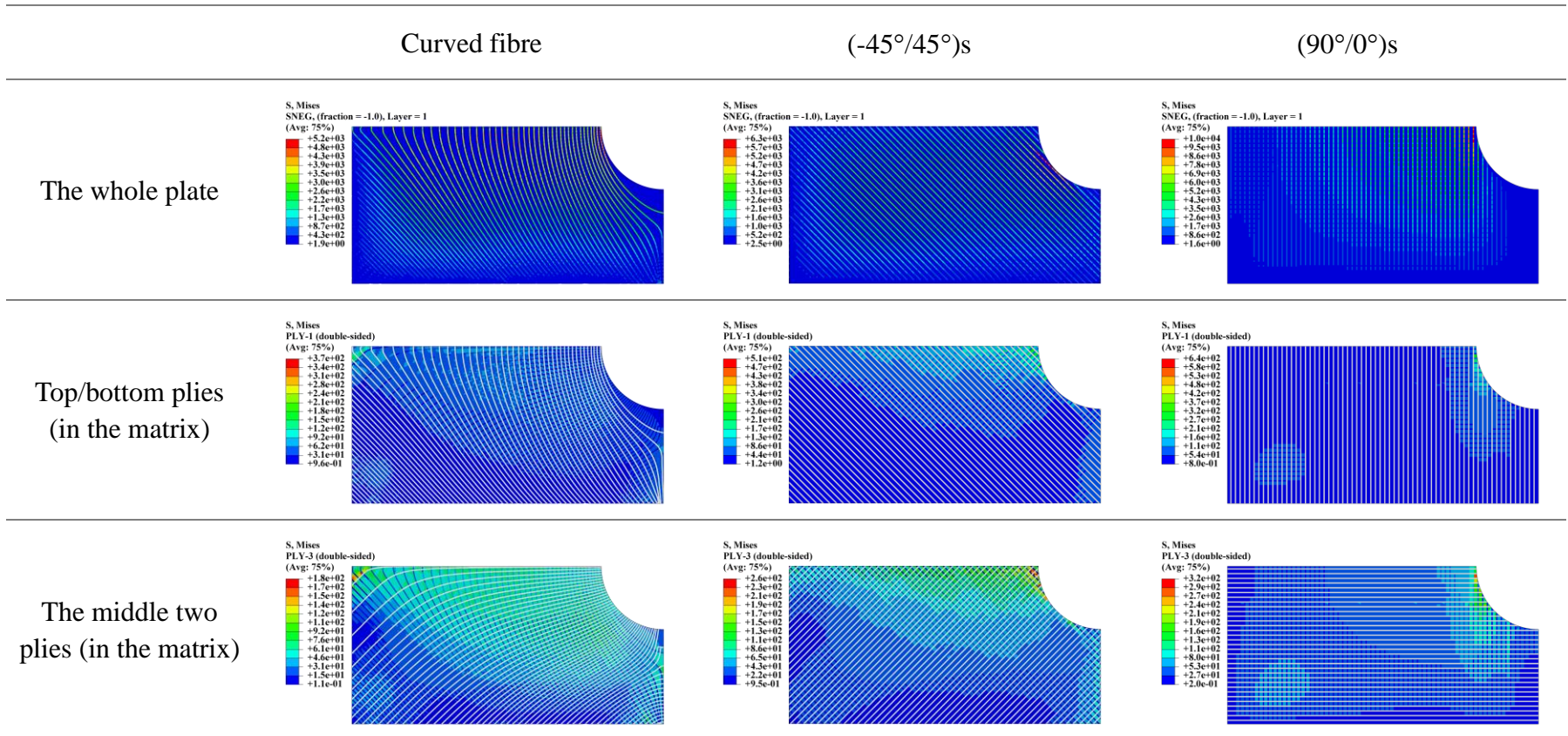

## 5. Conclusions

This paper presents a new optimization method to place continuous curved fibres without inter-section specifically for future application to 3D printing of CFRP composites. The method adopts maximum or middle principal stress trajectories as the placement path of carbon fibres, which aims to fully utilise the mechanical properties of fibres. By altering the number of placement path, the volume fraction of fibres in 3D printed composites can be controlled. The mechanical performance of optimised-printing CFRP composites are discussed in comparison with traditional optimization method with unidirectional fibres. Three representative cases, i.e. an open-hole single ply lamina under uniaxial tension and an open-hole laminate under biaxial tension and normal pressure were used to assess the newly proposed fibre placement method. In all case studies, it is confirmed that the curved fibre placement follow the principal stress trajectories will significantly improve the mechanical performance of the composites specimens in terms of stiffness and strength comparing to the traditional unidirectional fibre placement.

This new concept of performance-driven optimization method could offer a very useful and powerful tool for the design of future 3D printing process for fibre reinforced composites with complex and discontinuous geometry. It is worth to mention that although stress distributions near the hole are improved by curved fibre placement, there are still some stress concentration. Also when the overlap of fibres appear, the effect of optimization starts to decrease. Future researches are required to address these issues and also the implications to the challenges of the 3D printing technology need to be discussed with the manufacturing industry.